# SIMULATION OF THE TOTAL ENERGY OF A TRIATOMIC SILICON CLUSTER IN THE FIRST ORDER OF PERTURBATION THEORY

*V. P. Koshcheev[1], Yu. N. Shtanov[2]*

[1]*Moscow Aviation Institute (National Research University), Strela Branch, Moscow oblast, Zhukovskii, 140180 Russia*
ORCHID: https://orcid.org/0000-0002-0724-9760, email: koshcheev1@yandex.ru

[2]*Tumen Industrial University, Surgut Branch, Surgut, 628404 Russia*
ORCHID: https://orcid.org/0000-0002-5822-3368, email: yuran1987@mail.ru

As part of a new approach to calculating the total energy of a diatomic molecule (cluster), it is shown that in the first order of perturbation theory, the total energy of a triatomic cluster is equal to the sum of the total energies of the diatomic clusters (molecules). If all three aluminum atoms are in the ground state of Al(2p), then the quantum of energy of collective electron oscillations (plasmon energy) in each of the diatomic clusters (molecules) is $\hbar\omega_{el.} \approx 14$ eV (electron volts).

Using a computer experiment in [1-2], new types of ordered structures consisting of atomic silicon are investigated. The total energy of the interaction between atoms is calculated using the equation of density functional theory, the solution of which for the ground state coincides with the solution of the Hartree–Fock equation in the first order of perturbation theory [3]. In this paper, the total energy of a triatomic cluster will be calculated within the framework of a new approach [4,5] to calculating the total energy of a diatomic molecule in the first order of perturbation theory.

We will describe a triatomic molecule using the stationary Schrodinger equation

$$H\psi = E\psi. \tag{1}$$

The Hamiltonian of equation (1) is represented as

$$H = H^0 + U. \tag{2}$$

$$U = \frac{Z_1 Z_2 e^2}{|\mathbf{r}_1 - \mathbf{r}_2|} + \sum_{j_1=1}^{Z_1}\sum_{j_2=1}^{Z_2} \frac{e^2}{|\mathbf{r}_1 + \mathbf{r}_{1j_1} - \mathbf{r}_2 - \mathbf{r}_{2j_2}|} - \sum_{j_1=1}^{Z_1} \frac{Z_2 e^2}{|\mathbf{r}_1 + \mathbf{r}_{1j_1} - \mathbf{r}_2|} - \sum_{j_2=1}^{Z_2} \frac{Z_1 e^2}{|\mathbf{r}_1 - \mathbf{r}_2 - \mathbf{r}_{2j_2}|} +$$

$$+ \frac{Z_1 Z_3 e^2}{|\mathbf{r}_1 - \mathbf{r}_3|} + \sum_{j_1=1}^{Z_1}\sum_{j_3=1}^{Z_3} \frac{e^2}{|\mathbf{r}_1 + \mathbf{r}_{1j_1} - \mathbf{r}_3 - \mathbf{r}_{3j_3}|} - \sum_{j_1=1}^{Z_1} \frac{Z_3 e^2}{|\mathbf{r}_1 + \mathbf{r}_{1j_1} - \mathbf{r}_3|} - \sum_{j_3=1}^{Z_3} \frac{Z_1 e^2}{|\mathbf{r}_1 - \mathbf{r}_3 - \mathbf{r}_{3j_3}|} + \qquad (3)$$

$$+ \frac{Z_3 Z_2 e^2}{|\mathbf{r}_3 - \mathbf{r}_2|} + \sum_{j_3=1}^{Z_3}\sum_{j_2=1}^{Z_2} \frac{e^2}{|\mathbf{r}_3 + \mathbf{r}_{3j_3} - \mathbf{r}_2 - \mathbf{r}_{2j_2}|} - \sum_{j_3=1}^{Z_3} \frac{Z_2 e^2}{|\mathbf{r}_3 + \mathbf{r}_{3j_3} - \mathbf{r}_2|} - \sum_{j_2=1}^{Z_2} \frac{Z_3 e^2}{|\mathbf{r}_3 - \mathbf{r}_2 - \mathbf{r}_{2j_2}|},$$

where $U = U_{12} + U_{13} + U_{23}$ – the potential energy of the interaction of three atoms; $\mathbf{r}_1$, $\mathbf{r}_2$ и $\mathbf{r}_3$ – coordinates of the first, second and third atomic nuclei; $\mathbf{r}_1 + \mathbf{r}_{1j_1}$, $\mathbf{r}_2 + \mathbf{r}_{2j_2}$ и $\mathbf{r}_3 + \mathbf{r}_{3j_3}$ – coordinates $j_1$-го, $j_2$-го и $j_3$-го the electrons of the first, second, and third atoms, respectively; $\mathbf{r}_1 - \mathbf{r}_2 = \mathbf{r}_{12} + \delta\mathbf{r}_{12}, \mathbf{r}_1 - \mathbf{r}_3 = \mathbf{r}_{13} + \delta\mathbf{r}_{13}$, $\mathbf{r}_3 - \mathbf{r}_2 = \mathbf{r}_{32} + \delta\mathbf{r}_{32}$; The collective oscillations of atomic electrons and nuclei are described by a vector $\delta\mathbf{r}_{mn} = \delta\mathbf{r}_{mn}^{nucl.} + \delta\mathbf{r}_{mn}^{el.}$; $r_{mn} = |\mathbf{r}_{mn}|$ – the distance between $n$-th and $m$-th atoms in a cluster.

The solution of equation (1) with the Hamiltonian (2) will be sought using perturbation theory. $\psi = \psi^0 + \psi^1 + \ldots$ и $E = E^0 + E^1 + \ldots$. We will look for the potential energy (electronic terms) of a triatomic cluster in the first order of perturbation theory.

$$E^1 = \langle \psi^0 | U | \psi^0 \rangle, \qquad (4)$$

where are the angle brackets $\langle \ldots \rangle$ were introduced by Dirac [6].

The Hamiltonian $H^0$ let's imagine in the form $H^0 = H_1^0 + H_2^0 + H_3^0$, where $H_i^0$ – the Hamiltonian of the $i$-th atom; $i=1,2,3$.

Solving the Schrodinger equation $H^0\psi^0 = E^0\psi^0$, we will search in the form $\psi^0 = \psi_1^0\psi_2^0\psi_3^0$ and $E^0 = E_1^0 + E_2^0 + E_3^0$, where the Schrodinger equation for the $i$-th isolated atom has the form

$$H_i^0 \psi_i^0 = E_i^0 \psi_i^0, \qquad (5)$$

where $\psi_i^0 = \psi_i^0(\mathbf{r}_{i1}, \mathbf{r}_{i2}, \ldots, \mathbf{r}_{iZ_i})$.

Averaging over quantum fluctuations of the location of atomic electrons will be carried out using the method [7], which Bethe used to calculate the atomic form factor, and averaging over collective oscillations of atomic electrons and nuclei will be performed by the square of the modulus of the wave function of a harmonic oscillator in the ground state. The average values will be denoted by $\langle \psi_1^0 | U | \psi_1^0 \rangle = \langle ... \rangle_{e1}$, $\langle \psi_2^0 | U | \psi_2^0 \rangle = \langle ... \rangle_{e2}$, $\langle \psi_3^0 | U | \psi_3^0 \rangle = \langle ... \rangle_{e3}$ and $\langle \psi^0 | U | \psi^0 \rangle_{pl} = \langle ... \rangle_{pl}$.

Decompose the potential energy of interaction (3) into the Fourier integral

$$U = \int \frac{d^3\mathbf{k}}{(2\pi)^3} \frac{4\pi e^2}{k^2} \left[ Z_1 Z_2 + \frac{1}{2} \sum_{m \neq n=1}^{3} V_{mn}(k) \exp(i\mathbf{k}\delta\mathbf{r}_{mn}) \right] \exp(i\mathbf{k}\mathbf{r}_{mn}), \quad (6)$$

where $V_{12}(k) = \sum_{j_1=1}^{Z_1} \sum_{j_2=1}^{Z_2} \exp(i\mathbf{k}(\mathbf{r}_{1j_1} - \mathbf{r}_{2j_2})) - Z_2 \sum_{j_1=1}^{Z_1} \exp(i\mathbf{k}\mathbf{r}_{1j_1}) - Z_1 \sum_{j_2=1}^{Z_2} \exp(i\mathbf{k}\mathbf{r}_{2j_2})$.

Let's average the potential energy of the interaction of three atoms (6)

$$\langle U \rangle = \frac{1}{2} \sum_{m \neq n=1}^{3} \int \frac{d^3\mathbf{k}}{(2\pi)^3} \frac{4\pi e^2}{k^2} \left[ Z_m - F_m(k) \right]\left[ Z_n - F_n(k) \right] \exp\left[ i\mathbf{k}\mathbf{r}_{mn} - k^2 \Delta_{mn} \right], \quad (7)$$

where $\left\langle \sum_{j_m=1}^{Z_m} \exp(i\mathbf{k}\mathbf{r}_{mj_m}) \right\rangle_{em} = F_m(k)$; $F_m(0) = Z_m$ – atomic form factors of the first, second and third atoms; $\left\langle \exp\left( i\mathbf{k} \sum_{m \neq n=1}^{3} \delta\mathbf{r}_{mn} \right) \right\rangle_{pl} = \exp\left[ -k^2 \sum_{m \neq n=1}^{3} \Delta_{mn} \right]$; $\Delta_{mn} = \Delta_{nm}$; $2\Delta_{mn} = \sigma_{nucl.}^2 + \sigma_{el.}^2$ – the sum of the average squares of the amplitude of collective atomic nuclear $\sigma_{nucl.}^2 = \frac{\hbar}{2\omega_{nucl.}\mu_{nucl.}}$ and electronic $\sigma_{el.}^2 = \frac{\hbar}{2\omega_{el.}\mu_{el.}}$ vibrations per degree of freedom; elastic constant $\omega_{el.}^2 \mu_{el.} = \omega_{nucl.}^2 \mu_{nucl.} = U''_{mn}(r_{mn}^{\min})$ there is a value of the second derivative in the minimum potential energy between the $n$-th and $m$-th atoms in the cluster.

Since $\omega_{el.}^2 / \omega_{nucl.}^2 = \mu_{nucl.} / \mu_{el.} \gg 1$, then the quantum of energy of collective oscillations of electrons (plasmon energy) is equal to

$$\hbar\omega_{el.} \approx 4\Delta_{mn} U''_{mn}(r_{mn}^{\min}). \tag{8}$$

It is known that the wave function of a triatomic cluster must be antisymmetric with respect to the permutation of coordinates determining the locations of electrons. If the wave function of a triatomic cluster is chosen as the product of the wave functions of isolated atoms $\psi^0 = \psi_1^0 \psi_2^0 \psi_3^0$, then the exchange forces will be taken into account only for the electrons of isolated atoms.

Equation (7) for the potential energy of interaction without taking into account the exchange forces between the electrons of various isolated atoms is written as

$$\langle U \rangle = \frac{1}{2} \sum_{m \neq n=1}^{3} U(r_{mn}, \Delta_{mn});$$
$$U(r_{mn}, \Delta_{mn}) = Z_m Z_n e^2 \Phi(r_{mn}, \Delta_{mn})/r_{mn}, \tag{9}$$

where $\Phi(r_{mn}, \Delta_{mn}) = \frac{2}{\pi} \int_0^\infty \frac{1}{k} \left[1 - \frac{F_m(k)}{Z_m}\right]\left[1 - \frac{F_n(k)}{Z_n}\right] \sin(kr_{mn}) \exp\left[-k^2 \Delta_{mn}\right] dk$ – the function of shielding the potential energy of a diatomic cluster (molecule) [5].

By direct substitution, it can be shown that the potential energy shielding function is a solution to a diffusion type equation.

$$\frac{\partial \Phi(r_{mn}, \Delta_{mn})}{\partial \Delta_{mn}} = \frac{\partial^2 \Phi(r_{mn}, \Delta_{mn})}{\partial r_{mn}^2}. \tag{10}$$

It was shown in [5] that taking into account the collective oscillations of atomic electrons leads to a self-consistent system of equations

$$\begin{cases} \dfrac{\partial U(r_{mn}, \Delta_{mn})}{\partial r} = 0, \\ \dfrac{\partial^2 U(r_{mn}, \Delta)}{\partial r_{mn}^2} = \dfrac{const.}{\Delta_{mn}^2}, \end{cases} \tag{11}$$

where $U(r_m, \Delta_m)$ depends on $\Delta_m$ according to (8).

Using equation (10), we rewrite (11) as

$$\begin{cases} \dfrac{\partial U(r_{mn}, \Delta_{mn})}{\partial r_{mn}} = 0 \\ \dfrac{\partial}{\partial \Delta_{mn}} \left( U(r_{mn}, \Delta_{mn}) + \Delta \dfrac{\partial^2 U(r_{mn}, \Delta_{mn})}{\partial r_{mn}^2} \right) = 0 \end{cases}. \qquad (12)$$

As a result of solving the system of nonlinear equations (12), we obtain the values $\Delta_{mn}$ и $r_{mn}^{\min.}$.

It is known [3] that instead of the multiparticle Schrodinger equation (5), single–particle approximations are constructed based on the Hartree-Fock equations or the formalism of density functional theory. The atomic form factor is the Fourier component of the atomic electron distribution density

$$F_m(k) = \int n_m(\mathbf{r}) \exp(-i\mathbf{k}\mathbf{r}) d^3\mathbf{r}, \qquad (13)$$

where $n_m(\mathbf{r}) = 2|\psi_{1s}(\mathbf{r})|^2 + 2|\psi_{2s}(\mathbf{r})|^2 + 6|\psi_{2p}(\mathbf{r})|^2 + 2|\psi_{3s}(\mathbf{r})|^2 + 2|\psi_{3p}(\mathbf{r})|^2 -$ the electron density for a silicon atom Si.

Wave functions that approximate the solution of the Hartree-Fock equation for isolated silicon atoms, and energy $E_i^0$ presented in [8].

The condition for the applicability of the first-order correction of perturbation theory to the energy of a system in an undisturbed state has the form

$$|E^1| = |\langle U \rangle| \ll |E^0|. \qquad (14)$$

It is known [9] that the correction of the second order of smallness to the total energy is constructed using wave functions calculated with accuracy up to the first order of perturbation theory, that is, the mutual polarization of isolated atoms is taken into account. The correction of the second order of smallness to the total energy should be much smaller than the distance between the energy levels (electron terms) of a diatomic molecule [9].

To solve the problem of computer simulation of the total energy of the interaction of two atoms, the TotalEnergy program was developed [10, 11]. In the program, the atomic form factor is constructed using wave functions [8], which approximate solutions of the Hartree-Fock equation for isolated atoms. The program is written in the C++ programming language, and the graphical interface is built using the Qt library [12, 13]. Numerical integration using Clenshaw–Curtis quadratures and the QAGS algorithm, which is implemented in the GNU Scientific Library (GSL), is used for numerical calculation of expression (9) [14]. The program is cross-platform and has a GPL license.

The numerical solution of the system of equations (12) was performed using Richard Brent's algorithm [15]. The reliability of the numerical method was verified in [5] by comparing it with the analytical result for the shielded Coulomb potential of an isolated atom. The numerical solution of the system of equations (12) using wave functions [8], which approximate the solutions of the Hartree-Fock equation for isolated atoms, is shown in Fig.1 and Table 1 for various states of the silicon molecule. The results of calculating the total energy for various states of a diatomic cluster (molecule) of silicon are shown in Fig. 2. It can be seen that the total energy for a silicon molecule in the ground state is Si(3p)+Si(3p) in the first order of perturbation theory is the smallest $U(r_{12}^{min}) \approx 2.478$ eV at $(r_{12}^{min}/a_0) = 5.258$ and $\Delta_{12} \approx 0.462$ Å, $a_0 = 0.529$ Å. Figure 3 shows the results of calculating the second derivative of the total energy of a silicon molecule for the Si(3p)+Si(3p). It can be seen that the value of the second derivative at $(r_{12}^{min}/a_0) = 5.258$ equal $U''(r_{12}^{min}) \approx 7.534$ eV/Å$^2$. The value of the quantum energy of collective electron oscillations (plasmon energy) in a diatomic cluster (molecule) Si(3p)+Si(3p) is obtained using formula (8) $\hbar\omega_{el.} \approx 13.937$ eV. The plasmon energy in silicon is known to be 17 eV [16]. Thus, all the spectrometric parameters turn out to be values of the same order as the experimental data [17]. These spectrometric parameters were found as a result of solving a system of nonlinear equations (12). To find the minimum of the total energy of an n–atomic cluster, it is necessary to find a joint solution of n – systems of nonlinear equations (12) for diatomic clusters.

Additional materials to the article are available in [11].

Table 1. Values of the main parameters for calculating the plasmon energy.

| Cluster | $r_{12}^{min}/a_0$ | $\Delta_{12}, \text{Å}^2$ | $U(r_{12}^{min}, \Delta_{12}), \text{eV}$ | $\dfrac{\partial^2 U(r_{12}^{min}, \Delta_{12})}{\partial r^2}, \text{eV}/\text{Å}^2$ | $\hbar\omega_{el.}, \text{eV}$ |
|---|---|---|---|---|---|
| Si(3p)+Si(3p) | 5.258 | 0.462 | -2.478 | 7.534 | 13.937 |
| Si(1d)+Si(1d) | 5.425 | 0.492 | -2.260 | 6.447 | 12.685 |
| Si(1s)+Si(1s) | 5.764 | 0.557 | -1.871 | 4.708 | 10.497 |

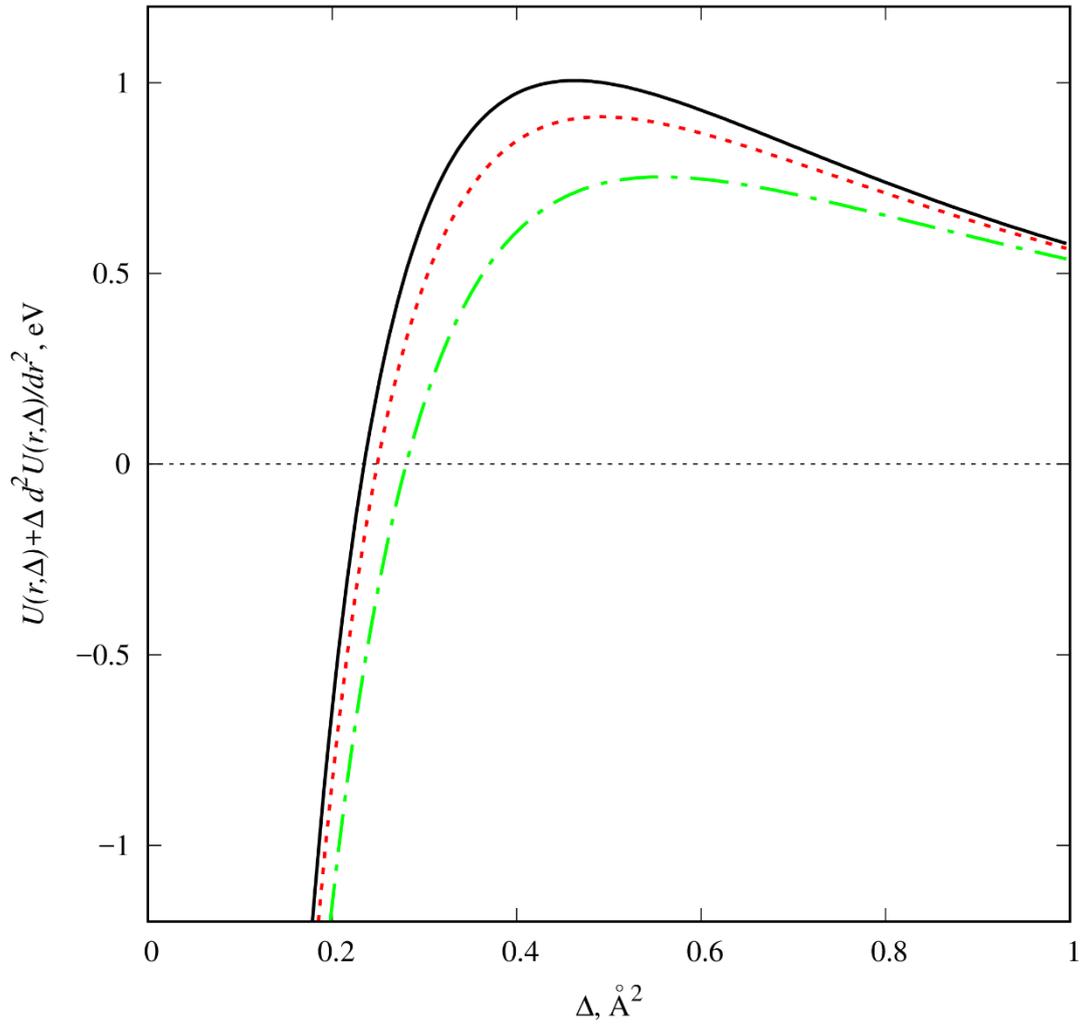

**Fig.1.** Graph of the function $U(r,\Delta) + \Delta \partial^2 U(r,\Delta)/\partial r^2$ at $\partial U(r,\Delta)/\partial r = 0$ for Si(3p)+Si(3p) (solid line), Si(1d)+Si(1d) (dotted line) and Si(1s)+Si(1s) (dashed dotted line).

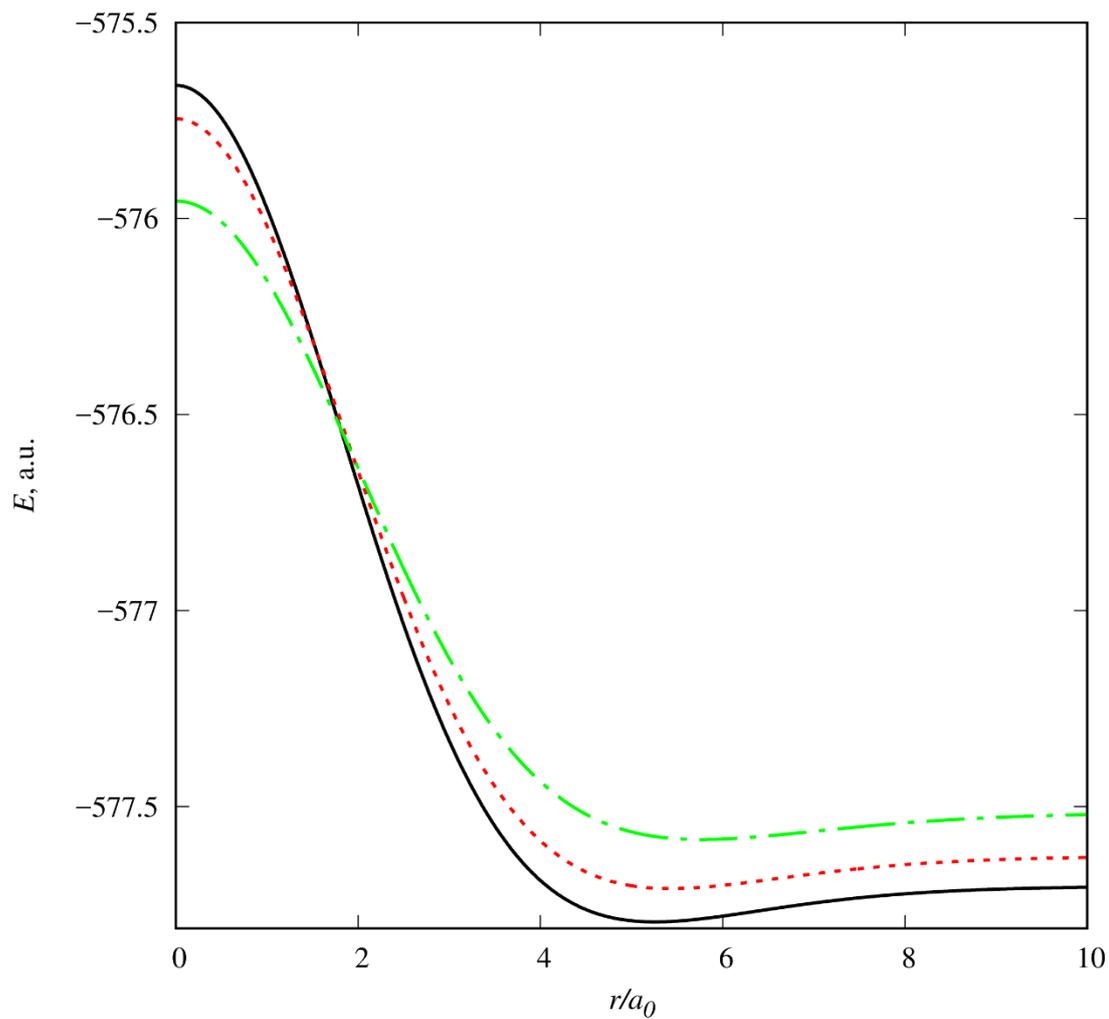

**Fig. 2.** The results of calculating the total energy of a silicon molecule for the states: solid line – Si(3$p$)+Si(3$p$); dotted line – Si(1$d$)+Si(1$d$); dashed dotted line – Si(1$s$)+Si(1$s$).

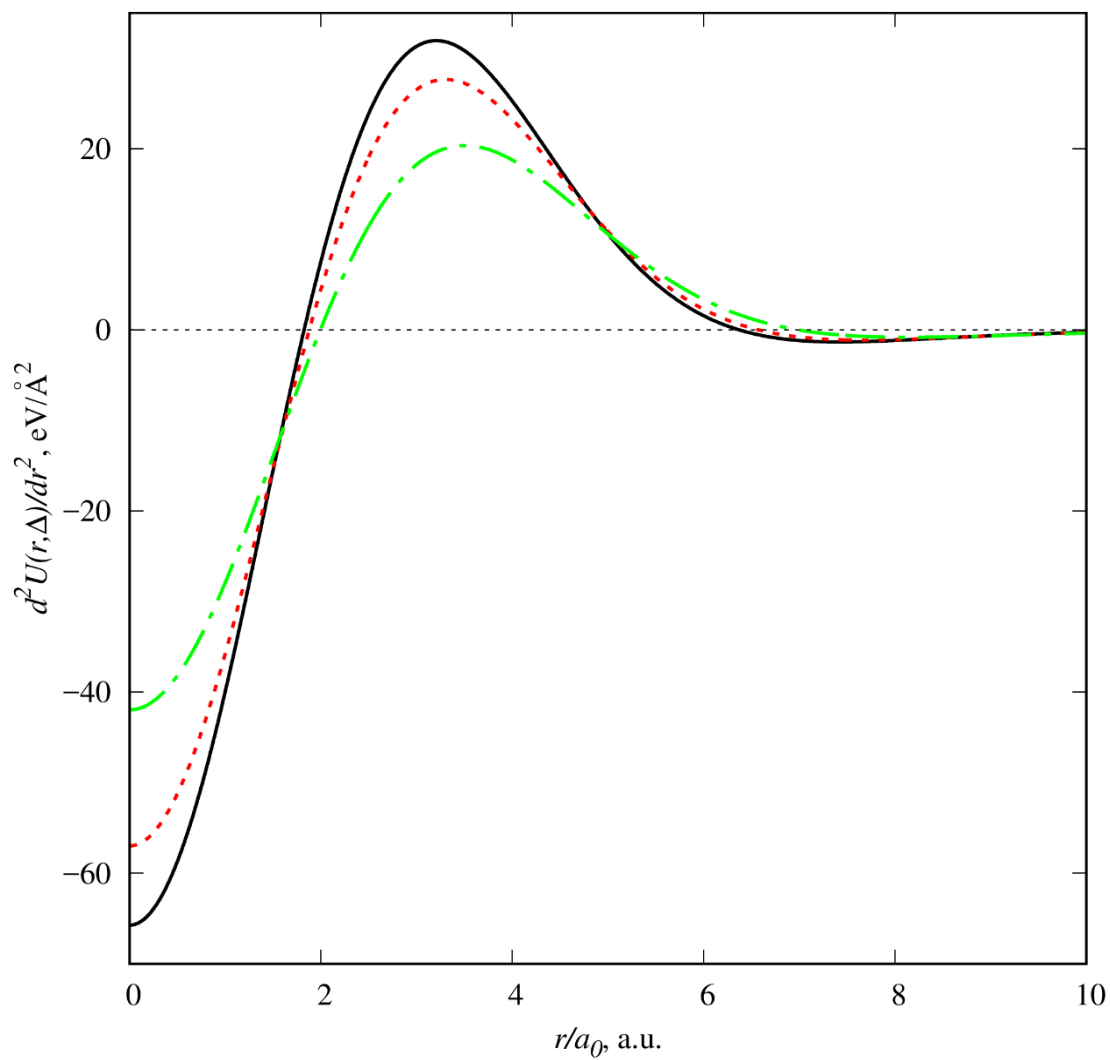

**Fig.3.** The results of calculating the second derivative of the total energy of a silicon molecule for the states: solid line – Si(3p)+Si(3p); dotted line – Si(1d)+Si(1d); dashed dotted line – Si(1s)+Si(1s).